\documentclass[conference]{IEEEtran}
\IEEEoverridecommandlockouts
\usepackage{cite}
\usepackage{amsmath,amssymb,amsfonts}
\usepackage{graphicx}
\usepackage{textcomp}
\usepackage{xcolor}
\def\BibTeX{{\rm B\kern-.05em{\sc i\kern-.025em b}\kern-.08em
    T\kern-.1667em\lower.7ex\hbox{E}\kern-.125emX}}

\usepackage{graphicx,wrapfig}
\usepackage{subcaption}
\usepackage{makecell}
\usepackage{multirow}
\usepackage{hyperref}

\usepackage{algorithm}
\usepackage{algpseudocode}

\usepackage{gensymb}
\usepackage{color}

\usepackage{accents}
\newcommand{\ubar}[1]{\underaccent{\bar}{#1}}

\graphicspath{{./Images/}}

\begin{document}


\title{Model Predictive Control based Energy Management System for Home Energy Resiliency\\
}



\author{\IEEEauthorblockN{1\textsuperscript{st} Ninad Gaikwad}
\IEEEauthorblockA{\textit{School of Electrical Engineering \& Computer Science} \\
\textit{Washington State University}\\
Pullman, USA \\
ninad.gaikwad@wsu.edu}
\and
\IEEEauthorblockN{2\textsuperscript{nd} Shishir Lamichhane}
\IEEEauthorblockA{\textit{School of Electrical Engineering \& Computer Science} \\
\textit{Washington State University}\\
Pullman, USA \\
shishir.lamichhane@wsu.edu}
\and
\IEEEauthorblockN{3\textsuperscript{rd} Anamika Dubey}
\IEEEauthorblockA{\textit{School of Electrical Engineering \& Computer Science} \\
\textit{Washington State University}\\
Pullman, USA \\
anamika.dubey@wsu.edu}
}

\author{\IEEEauthorblockN{Ninad Gaikwad, Shishir Lamichhane, and Anamika Dubey}
\IEEEauthorblockA{\textit{School of Electrical Engineering \& Computer Science} \\
\textit{Washington State University}\\
Pullman, USA \\
ninad.gaikwad@wsu.edu, shishir.lamichhane@wsu.edu, and anamika.dubey@wsu.edu}
}


\maketitle


\begin{abstract}
As the occurrence of extreme weather events is increasing so are the outages caused by them. During such unplanned outages, a house needs to be provided with an energy supply to maintain habitable conditions by maintaining thermal comfort and servicing at least critical loads. An energy system consisting of rooftop photovoltaic (PV) panels along with battery storage is an excellent carbon-free choice to provide energy resiliency to houses against extreme weather-related outages. However, to provide habitable conditions this energy system has to provide not only for the non-air-conditioning (non-AC) load demand but also for the turning on of the AC system which has a considerably higher startup power requirement as compared to its rated power. Hence, an intelligent automated decision-making controller is needed which can manage the trade-off between competing requirements.

In this paper, we propose such an intelligent controller based on Model Predictive Control (MPC). We compare its performance with a Baseline controller which is unintelligent, and a Rule-Based controller which has some intelligence, based on three resiliency metrics that we have developed. We perform extensive simulations for numerous scenarios involving different energy system sizes and AC startup power requirements. Every simulation is one week long and is carried out for a single-family detached house located in Florida in the aftermath of Hurricane Irma in 2017. The simulation results show that the MPC controller performs better than the other controllers in the more energy-constrained scenarios (smaller PV-battery size, larger AC startup power requirement) in providing both thermal comfort and servicing non-AC loads in a balanced manner. 
\end{abstract}






\section{Introduction}\label{sec:Introduction}
There has been a substantial increase in the occurrence of extreme weather events in recent years, where the average number of such events causing damages exceeding \$ 1 billion has increased from 8.1 each year to 22 events per year for the last three years (2021-2023)~\cite{disaster}. Hurricane Idalia (2023), the great Texas freeze (2021), California flooding (2023), Southern heatwaves (2023), and Hawaii wildfires (2023) are some well-known extreme weather events that occurred in the recent past that caused immense damages. In addition, such events can greatly damage the power systems infrastructure thereby causing long-unplanned power outages which lead to diverse socio-economic damages~\cite{USA_outages}. For example, the power system network of Puerto Rico and the US Virgin Islands experienced significant damages in 2017 as a result of Hurricanes Irma and Maria which led to the longest blackout, lasting almost 11 months leaving 3.3 million Puerto-Ricans without power~\cite{GAO_report}. 

A house equipped with rooftop solar PV together with battery storage systems (HPVB) to provide an excellent carbon-free alternative to provide electricity supply to residential customers during such an unplanned outage to maintain habitable conditions until the grid is restored~\cite{mango2021resilient}. Holistic habitable conditions can be maintained by providing some level of thermal comfort along with servicing the critical loads (refrigerators, freezers, cooking ranges, etc.) almost all the time, and servicing the non-critical loads whenever possible. Such exigencies for maintaining habitable conditions lead to a need for a Home Energy Management System (HEMS) that can manage the trade-off between competing requirements optimally.

A plethora of work has been done to develop HEMS for the grid-connected case (no outage) where the primary objective is to increase energy savings and/or decrease energy costs, see~\cite{review_demand_response1},~\cite{review_demand_response2},~\cite{review_not_off_grid}, and~\cite{review_HEMS}, but this is not the focus of our paper. On the contrary, the amount of work done to develop HEMS that can provide energy resiliency against extreme weather events is limited and can be broadly divided into two: planning-based approaches and operation-based approaches. The planning-based approaches in~\cite{stochastic_optimization_optimal_size_planning_battery}, and~\cite{chatterji2021planning} plan for optimal sizing of the PV and battery system so that the dual objectives of reducing the cost of the energy system and providing resiliency against outages are met. However, as these approaches do not provide any intelligent operational support during the outage there can arise situations where the planned system size is inadequate to provide for habitable conditions owing to a multitude of reasons like longer than expected outages, lower than expected solar irradiance, and changes in the expected house demand patterns. 

Now, to reliably provide energy resiliency to a house, operation-based approaches which provide explicit control decisions to optimally manage the home energy system during outages, have been developed to mitigate the shortcomings of the planning-based approaches. In our previous works~\cite{gaikwad2020smart}, and~\cite{raman2021reinforcement} we have developed operation-based HEMS that are based on MPC and Reinforcement Learning (RL) respectively for an HPVB; but, we have not considered AC operation, a multi-level load prioritization scheme, or real house load demand data. Moreover,~\cite{haessig2019resilience} has developed an MPC-based HEMS for an HPVB similar to our work~\cite{gaikwad2020smart}; but, it does not consider AC operation, any discrete on-off decisions, or a realistic load prioritization scheme. Also,~\cite{candan2023home} has developed a rule-based HEMS for an HPVB with an additional Electric Vehicle (EV) battery system; but, it does not consider AC operation, circuit-level load grouping, real house load demand data, or the effect of variation of PV and battery size. Finally~\cite{weienergy} is the closest to our work to the best of our knowledge, it has developed an HEMS based on MPC for an HPVB; but, it does not consider the higher startup power requirement of the AC, a multi-level load prioritization scheme, any discrete on-off decisions, the effect of variation of PV and battery size, or a multi-dimensional resiliency metric to evaluate the controller performance.

In this work, we develop an operation-based HEMS based on an MPC architecture to provide holistic home energy resilience against long-unplanned power outages caused by extreme weather events. The major contributions of this work: 1. We consider the operation of AC along with its higher startup power requirement (more realistic problem formulation), 2. We consider a flexible multi-level load prioritization scheme that is based on circuit-level load grouping (easier real-world implementation, see the products and services/products of \href{https://www.span.io/}{SPAN.io, Inc.}), 3. We consider the discrete AC on-off decisions explicitly in our optimization formulation for the MPC, while the on-off decisions for the circuit-level prioritized loads are computed by a separate priority stack controller (making the resulting mixed integer linear program computationally simpler), 4. We assess the performance of our proposed controller against two other controllers based on a three-dimensional resiliency metric, 5. We consider the effect of different PV, and battery system sizes and AC startup power requirements while assessing the performance of our proposed controller, and 6. We use real house load demands, and weather data after a real extreme weather event for our simulations.  

The rest of the paper is organized as follows: the system is described in Section~\ref{sec:SystemDescription}, the plant model that the controllers try to manage is explained in Section~\ref{sec:PlantModel}, the mathematical formulations for the proposed MPC controller along with the baseline controller and rule-based smart controller are laid out in Section~\ref{sec:ControlAlgorithms} the simulation setup used to generate results is discussed in Section~\ref{sec:SimulationSetup}, the simulation results are presented and discussed in Section~\ref{sec:ResultsDiscussion}, and the main conclusions are provided in Section~\ref{sec:Conclusion}.



\section{System Description}\label{sec:SystemDescription}
\begin{figure}[htpb]
	\centering
	\includegraphics[scale=0.21]{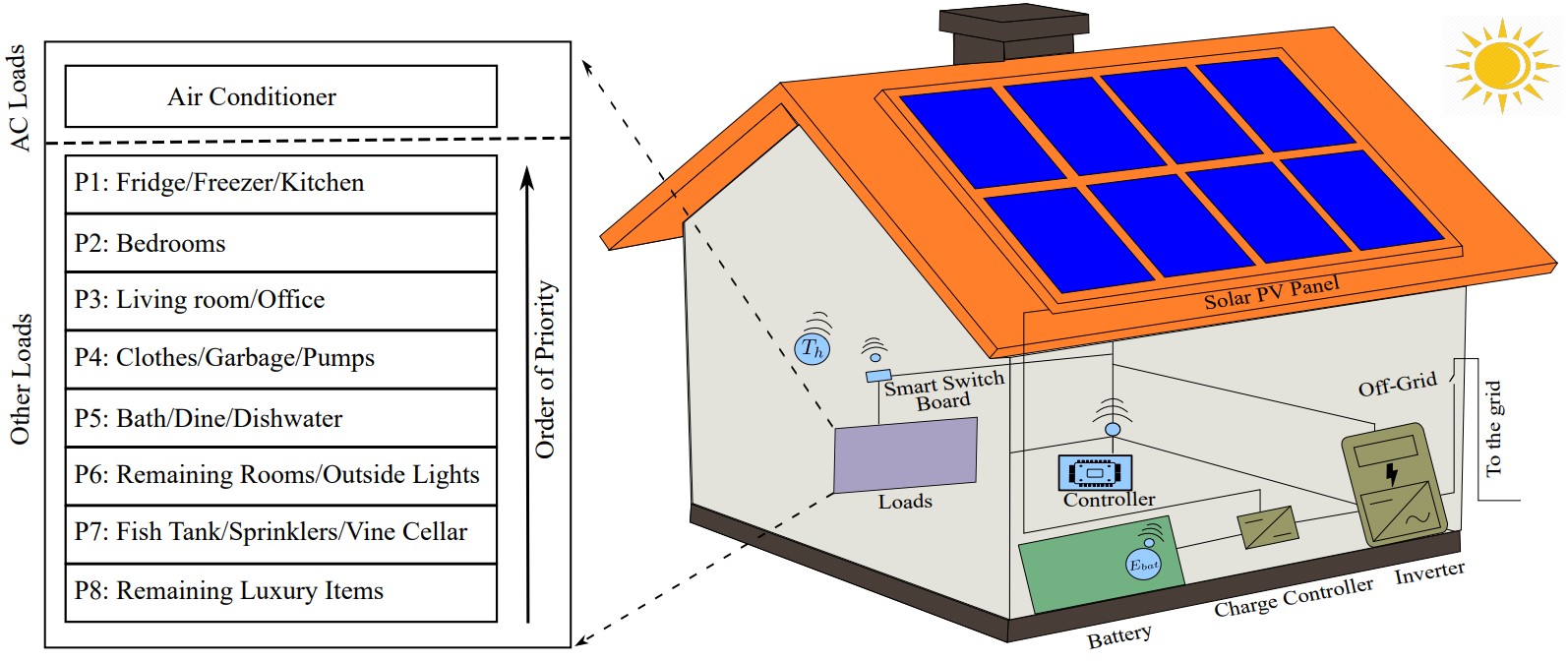}
	\caption{Schematic of the home energy system with its load description.}
	\label{fig:PlantSchematic}
\end{figure}
The Figure~\ref{fig:PlantSchematic} illustrates the home energy system with a detailed classification of its intrinsic loads (energy-consuming systems). The house is equipped with a solar PV system for energy generation and a battery system for energy storage. The two broad categories of the loads in the house are the AC load and non-AC loads referred to as other loads. The other loads are further categorized based on their priority level which depends on their subjective importance for maintaining a habitable condition for a given house based on the owner's preferences during a grid outage (we consider that grid outage is caused by an extreme weather event and it is not known when the grid will be restored). It should be noted that the priority levels and the loads that are part of it as shown in Figure~\ref{fig:PlantSchematic} are not absolute but a preference of the authors. The objective of any well-designed controller for this system will be to serve as much load as possible to maintain maximum comfort conditions with the constraint of finite energy by controlling the charging/discharging of the battery system, the on-off of the AC system, and the on-off of the prioritized other loads. Furthermore, two sensors are present to measure the internal house temperature and the battery energy level so that controllers can use these measurements as feedback if and when required. Also, we consider that the on-off of the AC system and that of the prioritized other loads are realized through a smart switchboard (each circuit corresponding to both the AC load and individual other loads has a smart switch associated with them in the smart switchboard). 


\section{Plant Model}\label{sec:PlantModel}
 The plant model is a mathematical description of the real system, using which a digital simulator can be developed for testing the performance of a controller (being designed for managing the real system) in a closed-loop simulation. Now for describing our system (presented in Section~\ref{sec:SystemDescription}) mathematically we consider time to be discrete, $k = 0,1,2,3,\dots,$ denotes the time index, and $\Delta T_s$ is the time between any two adjacent time indices (i.e time-step). Moreover, $E(k)$ is any energy produced or consumed by any subsystem between the time interval marked by time indices $k-1$ and $k$. The dependence on $k$ for any variable can be omitted where necessary and constants will be mentioned explicitly. Any variables with a bar on top ($\bar{.}$) or bar on the bottom ($\ubar{.}$) are either constants or data (i.e. not computed but provided). The units for the upcoming energy, power, and temperature variables are $kWh$, $kW$, and $^{\circ}C$ respectively.   
  
 In our system, the maximum energy available from the PV system ($\bar E_{pv}$) is modeled using a set of equations given in \cite{gaikwad2020smart} and \cite{faiman2008assessing}, and it is affected by weather variables: solar irradiance ($\bar GHI \; (kW/m^{2})$), ambient outside temperature ($\bar T_{am}$), and wind speed ($\bar W_{s} \; (m/s)$). In addition, it has two dynamic systems: the house thermal dynamics system that gives the time evolution of $T_{h}$ (internal house temperature), and the battery energy dynamics system that gives the time evolution of $E_{bat}$ (battery energy level). These are modeled as linear discrete-time dynamical systems and are presented in detail in \cite{cui2019hybrid} and \cite{gaikwad2020smart} respectively. The control commands for our system are given as follows: battery system control command is $u_{bat} = [c, d]^{T}$ where $c, d \in \{0,1\}$ (binary variables) are the battery charging and discharging commands respectively, AC system on-off control command is $u_{ac} \in \{0,1\}$ (binary variable), and other loads on-off control commands are given as $u_{l_{i}} \in \{ 0, 1\} \; \forall i \in I = \{1,\dots,n \}$ (binary variables) where $n$ is the number of different prioritized loads and the set $I$ is ordered in descending order of priority. The energy consumed by AC now is given as $u_{ac} \bar E_{ac}$ where $\bar E_{ac}$ is rated energy consumption of the AC system, and that consumed by the other loads is given by $E_{l} = \sum_{i=1}^{n} u_{l_{i}} \bar E_{l_{i}}$ and we have $\bar E_{l} = \sum_{i=1}^{n} \bar E_{l_{i}}$ where $\bar E_{l_{i}}$ is the energy demanded by the $i^{\text{th}}$ priority load and $\bar E_{l}$ is total energy demand from the other loads ($i \in I$). The remaining plant model consists of equations representing the energy consumption/production interaction between the Battery system, PV system, AC system, and other loads. These equations are similar to the ones described in \cite{gaikwad2020smart} with an additional power mismatch constraint over the energy mismatch constraint (both are explained in Section~\ref{subsec:SmartController}).


\section{Control Algorithms and Performance Metrics}\label{sec:ControlAlgorithms}
\begin{figure*}[!t]
	\centering
	\begin{subfigure}[t]{0.48\textwidth}
		\centering
		\includegraphics[scale=0.22]{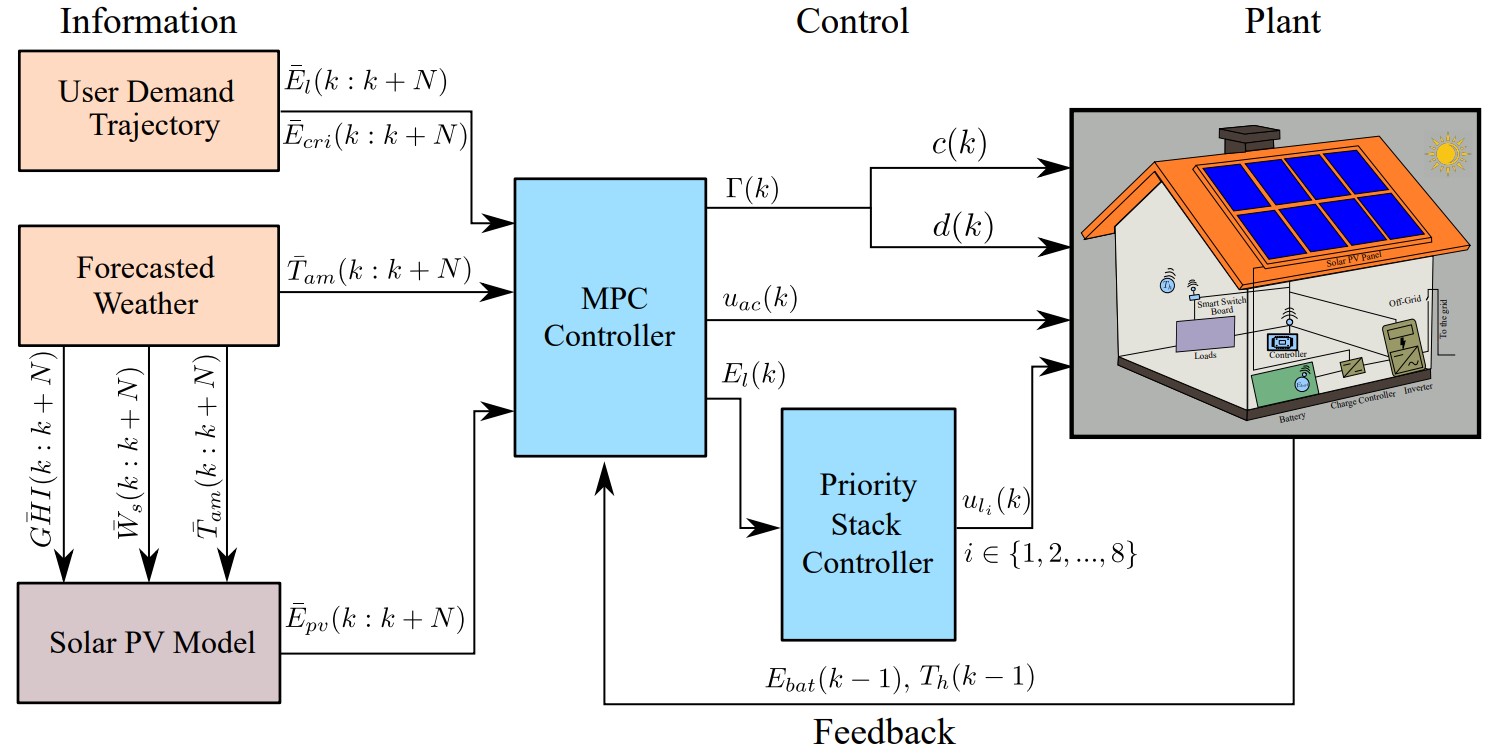}
		\caption{MPC Controller Schematic.}
		\label{fig:MPCSchematic}
	\end{subfigure}
	\begin{subfigure}[t]{0.48\textwidth}
		\centering
		\includegraphics[scale=0.30]{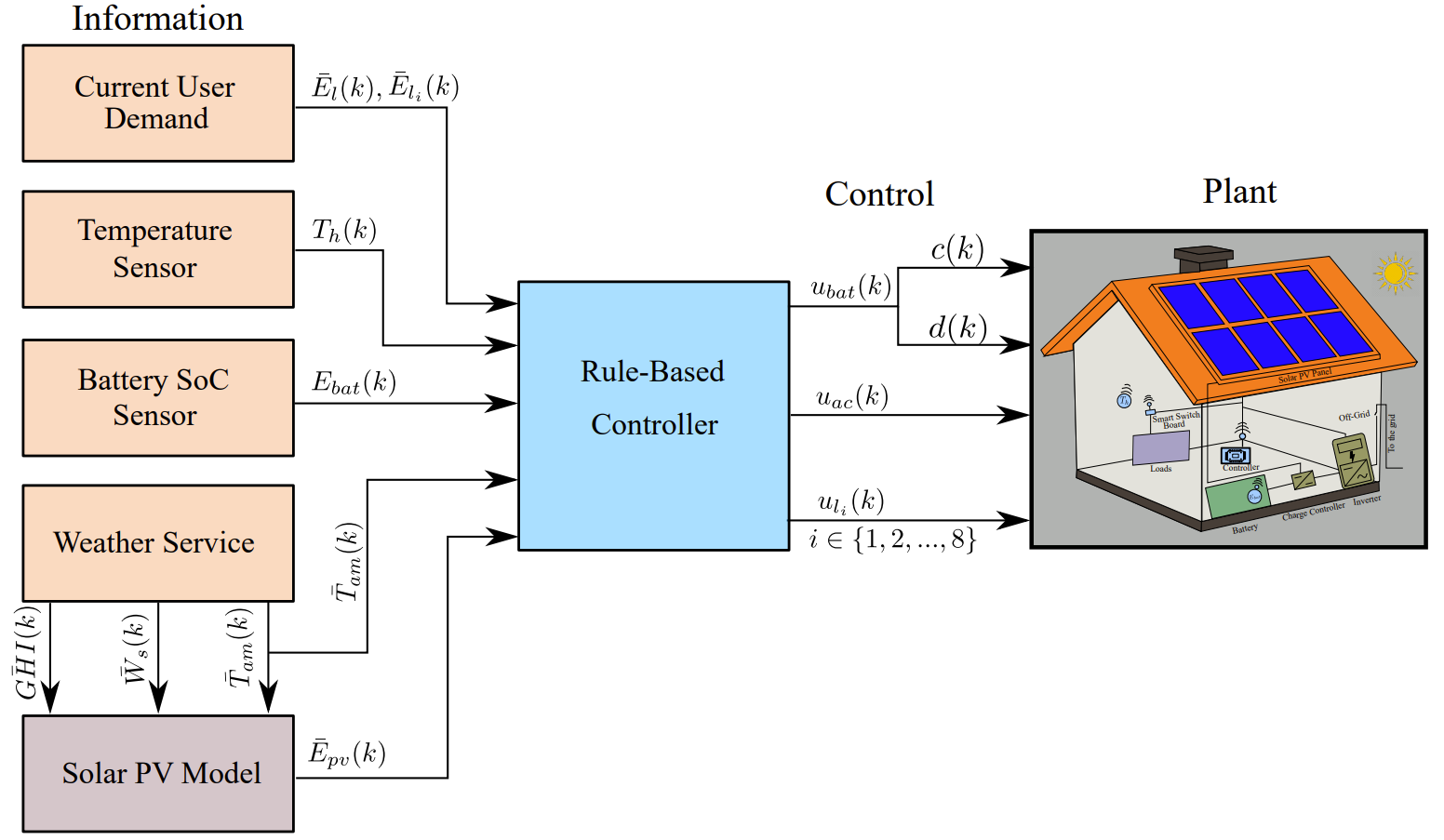}
		\caption{Rule-Based Smart Controller Schematic.}
		\label{fig:RuleBasedSchematic}
	\end{subfigure} 
	\caption{Schematics of MPC and Rule-Based Smart Controllers..}
	\label{fig:MPCRuleBasedSchematic}
\end{figure*}

\subsection{Model Predictive Controller}\label{subsec:MPCController}
The MPC illustrated in Figure~\ref{fig:MPCSchematic} computes the control commands at discrete time indices $k = 1,2 \dots N$ where $N$ is the total number of time indices in the planning horizon (only the first time index i.e the $j^{\text{th}}$ time index control commands are sent to the plant), for the next time index $j+1$, the computation is done taking into account the feedback of the dynamic states of the plant and the new available forecasts/estimates of the disturbances). The decision variables for the optimization problem elemental to the MPC controller are as follows: the states of the process $x(k) = [T_{h}(k), E_{bat}(k)]^{T}$; the control commands $u(k) = [\Gamma(k),u_{ac}(k), E_{l}(k)]^{T}$, where $\Gamma(k)$ (converted to realizable control commands as shown in (\ref{eq:BatteryCharging}) and (\ref{eq:BatteryDischarging})), $u_{ac}(k)$ and $E_{l}(k)$ (converted to realizable control commands by the Priority Stack Controller described in Section~\ref{subsubsec:PriorityStackController}) are the fraction of the normal battery charging energy, the AC on-off control command and energy consumed by the other loads respectively (it should be noted that having $\Gamma(k) \in \mathbb{R}$ eliminates 2 binary variables, while having $E_{l}(k)$ eliminates $n$ binary variables); the internal variables $v(k) = [E_{pv}(k),\zeta_{h}(k), \zeta_{l}(k), f_{on}(k), f_{off}(k), \theta_{bat}(k)]^{T}$, where $E_{pv}(k)$ is the energy produced by the PV panels, $\zeta_{h}(k)$ and $\zeta_{l}(k)$ are the slack variable for house temperature and minimum critical load ($\bar E_{cri} \triangleq \bar E_{l_{1}}$ i.e. the highest priority load) respectively to ensure feasibility, $f_{on}(k)$ and $f_{off}(k)$ are the AC turn-on and turn-off indicators as described in Eq.(\ref{eq:ACFlipOn}) and Eq.(\ref{eq:ACFlipOff}), and $\theta_{bat}(k)$ is the battery discharge indicator. The exogenous inputs whose predictions are assumed to be known for the $N$ time step $w(k) = [\bar E_{pv}(k),\bar E_{l}(k), \bar T_{am}(k)]^{T}$, where $\bar E_{pv}$, $\bar E_{l}(k)$, and $\bar T_{am}$ are the available energy from the PV panels, the forecast of the total other load demand of the house, and the ambient outside air temperature respectively. Hence, the complete decision vector for the optimization problem is given as $[X,U,V]^{T}$, where $X:=[x(k+1), \dots,x(k+N)]^{T}$, $U:=[u(k), \dots , u(k+N-1)]^{T}$ and $V:=[v(k), \dots , v(k+N-1)]^{T}$. Moreover; $u_{ac}$, $f_{on}$, $f_{off}$ and $\theta_{bat}$ are binary variables i.e. $\in {0,1}$ while rest all variables $\in \mathbb{R}$.

\begin{align}
    c(k) &=  
	\begin{cases}
		1 \; , & \text{if } \Gamma(k) > 0  , \\
		0 \; , & \text{if } \Gamma(k) \leq 0  , \\
	\end{cases} \label{eq:BatteryCharging}\\
	d(k) &= 
	\begin{cases}
		1 \; , & \text{if } \Gamma(k) < 0   ,\\
		0 \; , & \text{if } \Gamma(k) \geq 0  , \\
	\end{cases} \label{eq:BatteryDischarging}\\
	f_{on}(k) &=  \label{eq:ACFlipOn}
	\begin{cases}
		1 \; , & u_{ac}(k) = 1 \; \&  \; u_{ac}(k-1) = 0  , \\
		0 \; , & \text{otherwise,} 
	\end{cases} \\ 
	f_{off}(k) &=  \label{eq:ACFlipOff}
	\begin{cases}
		1 \; , & u_{ac}(k) = 0 \; \& \; u_{ac}(k-1) = 1  , \\
		0 \; , & \text{otherwise.} 
	\end{cases} 	
\end{align}

\subsubsection{Optimization Problem Formulation}\label{subsubsec:OptimizationProblemFormulation}
The MPC solves a Mixed Integer Linear Programming (MILP) problem to compute the control commands over the planning horizon, and an instance of this problem at any time index $j$ is given mathematically as follows:
\begin{equation}
	\begin{aligned} 
		\min_{X,U,V} & \sum\limits^{j+N-1}_{k=j} \bigg[\lambda_{1}(N-k)  \zeta_{h}(k) + \lambda_{2}(N-k)  \zeta_{l}(k) - \\
		&\quad \quad \lambda_{3}(N-k)  E_{l}(k) - \lambda_{4}  E_{bat}(k) + \\	
		&\quad \quad \lambda_{5}  \theta_{bat}(k) \bigg], \label{eq:CostFunction_SingleHouse}  
	\end{aligned}
\end{equation}
subject to the following constraints:
\begin{align}
	&T_{h}(k+1)=AT_{h}(k) + Bu_{ac}(k) Q_{ac} + D \bar T_{am}(k), \nonumber \\
	 \label{eq:House_eqcon_1}  \\
	&E_{bat}(k+1) = E_{bat}(k) - \Gamma(k)   \bar E_{bat}^{c,dc}, \nonumber \\ 
	 \label{eq:Battery_eqcon_1}\\
	&  u_{ac}(k) \bar E_{ac}-  \Gamma(k) \bar E_{bat}^{c,dc} +  E_{l}(k)=   E_{pv}(k), \label{eq:EnergyBalance_eqcon_1} \\		
	& u_{ac}(k) = \sum\limits^{k}_{m=j} f_{on}(m) - \sum\limits^{k}_{m=j} f_{off}(m), \label{eq:ACControl_eqcon_1} \\
	&  f_{on}(k) \bar P_{ac}  \leq  \theta_{bat}(k) \bar P_{bat}^{dc} +  \frac{\bar E_{pv}(k)}{\Delta T_{s}}, \label{eq:ACControl_eqcon_2}  \\
	&\ubar T_{h}\leq T_{h}(k)\leq \bar T_{h} + \zeta_{h}(k),  \label{eq:House_ineq_1} \\	
	&\ubar E_{bat}\leq E_{bat}(k)\leq \bar E_{bat},   \label{eq:Battery_ineq_1} \\
	& \ubar\Gamma \leq \Gamma(k) \leq \bar\Gamma,   \label{eq:Battery_ineq_2} \\
	&\bar E_{cri}(k) - \zeta_{l}(k)  \leq E_{l}(k) \leq \bar E_{l}(k),  \label{eq:Load_ineq_1} \\	
	&0 \leq E_{pv}(k) \leq \bar E_{pv}(k),  \label{eq:PV_ineq_1} \\
	&  f_{on}(k) + f_{off}(k) \leq 1, \label{eq:ACControl_eqcon_3} \\    
	&\zeta_{h}(k)\geq 0, \label{eq:Slack_ineq_1}  \\
	&\theta_{bat}(k)\geq \Gamma(k),  \label{eq:Slack_ineq_2} \\
	&\bar E_{cri}(k) \geq \zeta_{l}(k)\geq 0.  \label{eq:Slack_ineq_3}      
\end{align}

\subsubsection{Cost Function Description}\label{subsubsec:CostFunctionDescription}
 The cost function is given in (\ref{eq:CostFunction_SingleHouse}), 
 where the first term in the cost function tries to minimize the temperature slack variable $\zeta_{h}$, which implicitly tries to keep the house temperature within bounds, see (\ref{eq:House_ineq_1}). The second term tries to minimize energy demand slack variable $\zeta_{l}$, which implicitly tries to maximize the energy provision for critical load demand, see (\ref{eq:Load_ineq_1}). The third term tries to maximize the house energy demand variable $E_{h}$ (excluding AC energy demand) as we would like most of the house energy demand to be met most of the time. The fourth term tries to maximize the battery energy level $E_{bat}$, this is to ensure that the battery is charged when there is excess solar and that it is maintained at an appropriate level to provide for future house energy demand. The final term tries to minimize the battery charging mode indicator $\theta_{bat}$, this minimization coupled with the constraint in (\ref{eq:Slack_ineq_2}) helps carve out the right behavior of this variable i.e. it takes the value of 1 when the battery is in charging mode and take the value 0 when the battery is in discharging mode. The $\lambda$s are weighting factors for the cost function terms to achieve the right trade-off between the competing objectives resulting in desired control behavior. The $(N-k)$ terms present with certain cost function terms are also additional weighting factors to incorporate temporal weighting which descends through the planning horizon. The particular objective with this temporal weighting is desired to be achieved for the present and near future but is deemed okay to fail in the far future w.r.t. the planning horizon, as this is a finite energy system we desire certain of our objectives to be achieved myopically/greedily w.r.t. time.

\subsubsection{Constraints Description}\label{subsubsec:ConstraintsDescription}
 (\ref{eq:House_eqcon_1}) and (\ref{eq:Battery_eqcon_1}) are the house thermal dynamics and the battery dynamics respectively (there is no plant-model mismatch for house thermal dynamics while there is a plant-model mismatch for battery dynamics as we have modeled $\Gamma$ instead of $c$ and $d$ and ignored the charging/discharging efficiencies). (\ref{eq:EnergyBalance_eqcon_1}) is the energy balance equation (there is a plant-model mismatch for energy balance as we have modeled $E_{l}$ instead of $\sum_{i=1}^{n}u_{l_{i}} \bar E_{l_{i}}$). (\ref{eq:ACControl_eqcon_1}) is the relationship of the AC on-off control variable with the binary indicators for the turning on ($f_{on}$) and turning off ($f_{off}$) of the AC. (\ref{eq:ACControl_eqcon_2}) conveys the AC startup power constraint where $f_{on}$ indicates the turning on of AC while the binary indicator $\theta_{bat}$ indicates the charging mode of the battery. (\ref{eq:House_ineq_1}) gives the bounds (upper bound $\bar T_{h}$ and lower bound $\ubar T_{h}$) for the house temperature where $\zeta_{h}$ serves as the slack variable for the upper bound of the house temperature (we require it due to finite energy supply of the system). (\ref{eq:Battery_ineq_1}) gives the bounds (upper bound $\bar E_{bat}$ and lower bound $\ubar E_{bat}$) on the battery energy storage, while (\ref{eq:Battery_ineq_2}) gives the bounds (upper bound $\bar \Gamma$ and lower bound $\ubar \Gamma$) on the charging/discharging rate of the battery. (\ref{eq:Load_ineq_1}) provides upper and lower bounds for the load energy demand variable where $E_{cri}$ is the energy demanded by the critical loads, $\bar E_{l}$ is the actual total energy demand of the house (AC excluded), and $\zeta_{l}$ is the slack variable (we require it due to finite energy supply of the system).(\ref{eq:PV_ineq_1}) gives the bound on the PV energy production where $\bar E_{pv}$ is the maximum energy production potential of the installed PV. (\ref{eq:ACControl_eqcon_3}) shows the complementary nature of the AC turning on ($f_{on}$) and turning off ($f_{off}$) binary indicators. (\ref{eq:Slack_ineq_1}) and (\ref{eq:Slack_ineq_3}) provide for the slack variables $\zeta_{h}$ and $\zeta_{l}$ to be positive respectively. (\ref{eq:Slack_ineq_2}) pits the battery charging mode indicator $\theta_{bat}$ against the battery charging/discharging rate so that it takes the value of 1 only when the battery is in the charging mode (we have added minimization of $\theta_{bat}$ to the cost function for appropriate behavior).

\subsubsection{Priority Stack Controller}\label{subsubsec:PriorityStackController}
The priority stack controller converts the first time index ($j$ w.r.t. the MILP planning horizon) control command value $E_{l}(j)$ into realizable control commands $u_{l_{m}}(j) \; , m \in I$ which can be sent to the plant. The priority stack controller logic is given as follows;
\begin{align}
    u_{l_{m}}(j) &=  
	\begin{cases}
		1 \; , & \text{if } \sum_{i=1}^{m} \bar E_{l_{m}}(j) \leq E_{l}(j) ,\\
		0 \; , & \text{otherwise,} \\
	\end{cases} \label{eq:PriorityStackLogic}  
\end{align}
and the complete control command for the priority stack controller can be given as $u_{l}^{ps} = [ u_{l_{1}},\dots,u_{l_{m}},\dots,u_{l_{n}} ]^{T}$.

\subsection{Baseline Controller}\label{subsec:BaselineController}
\begin{figure}[htpb]
	\centering
	\includegraphics[scale=0.21]{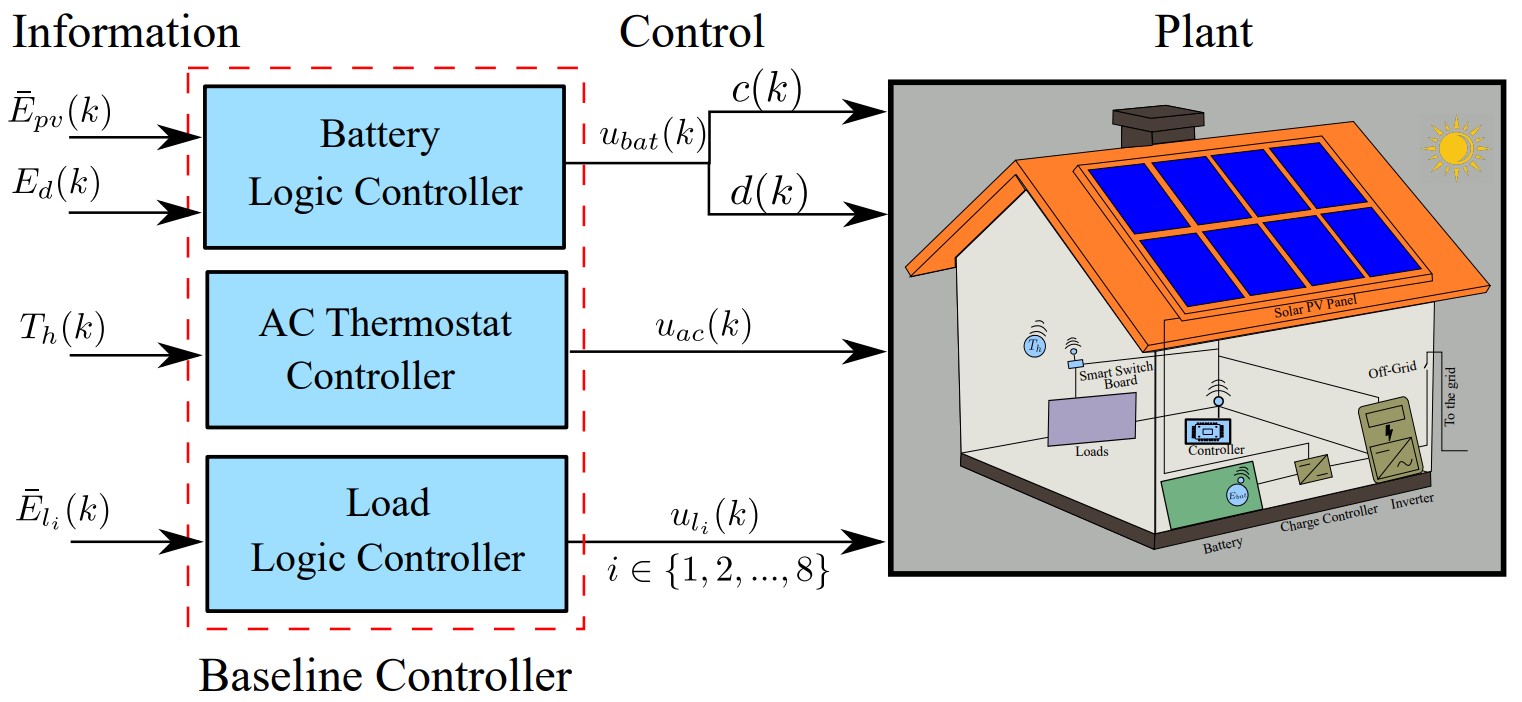}
	\caption{Schematic of the Baseline Controller.}
	\label{fig:BaselineSchematic}
\end{figure}
The baseline controller illustrated in Figure~\ref{fig:BaselineSchematic} is meant to abstract the operation of a commercially installed PV-Battery system for a house in the absence of any intelligent/supervisory control during a grid outage. It consists of three controllers: the battery control logic, the AC thermostat controller, and the load control logic. It should be noted these three controllers operate independently of each other at the level of their concerned subsystem, and that a smart switchboard is not present.    

\subsubsection{Battery Control Logic}\label{subsubsec:BatteryControlLogic}
The battery control logic commands the charging/discharging of the battery and is given as follows: 
\begin{align}
c(k)= 
\begin{cases}
1,& \text{if } \bar E_{pv}(k) \geq E_{d}(k) \\
0,& \text{otherwise } \\
\end{cases} \label{eq:Baseline_Charging} \\
d(k)= 
\begin{cases}
1,& \text{if } E_{pv}(k) < E_{d}(k) \\
0,& \text{otherwise,} \\
\end{cases} \label{eq:Baseline_Discharging} 
\end{align}
where $E_{d}$ is the total house energy demand given by $E_{d}(k) = E_{l}(k) + u_{ac}(k) \bar E_{ac}$. This controller is a mathematical abstraction of the interaction of the battery system with the rest of the subsystems in the house along with the logic present in the charge controller which connects the battery to the inverter (the battery charging/discharging is not controlled by an external controller but rather is a manifestation from the intrinsic logic and circuit conditions).

\subsubsection{AC Thermostat Controller}\label{subsubsec:ACThermostatController}
The AC thermostat controller commands the on-off of the AC and is given as follows: 
\begin{align}
	u_{ac}(k)= 
	\begin{cases}
		1,& \text{if }  T_{h}(k) \geq \bar T_{h}\\
		0,& \text{if }  T_{h}(k) \leq \ubar T_{h}\\
		u_{ac}(k-1),& \text{otherwise.}
	\end{cases} \label{eq:ACThermostatController} 
\end{align}
This is a realized controller that is intrinsic to the AC system. 

\subsubsection{Load Control Logic}\label{subsubsec:ALoad Control Logic}
The load logic controller controls the on-off of the other loads and is given as follows:
\begin{align}
	u_{l_{i}}(k)= 
	\begin{cases}
		1,& \text{if }  \bar E_{l_{i}}(k) > 0\\
		0,& \text{if }  \bar E_{l_{i}}(k) = 0.
	\end{cases} \label{eq:LoadLogicControl} 
\end{align}
This controller is a mathematical abstraction of the physical turning on and turning off of the individual other loads by any occupant of the house.

\subsection{Rule-Based Controller}\label{subsec:SmartController}
\begin{algorithm}[hbt!]
\caption{Algorithm for the Rule-Based Controller}\label{alg:RuleBasedController}
\begin{algorithmic}
\State Use the Solar PV model to compute $\bar E_{pv}(k)$
\State Use the Baseline Controller (\ref{eq:Baseline_Charging})-(\ref{eq:LoadLogicControl}) to compute $u^{b}(k)$.
\State Compute $E_{bat_{d}}(k)$ as given in (\ref{eq:BatteryDispatchDC}).
\State Compute $S_{ac}^{on}(k)$.
\State Compute $E_{mis}(k)$ using (\ref{eq:Emis}).
\State Compute $P_{mis}(k)$ using (\ref{eq:Pmis}).
\If{($E_{mis}(k) \ge 0$) \text{\textbf{AND}} ($P_{mis}(k) \ge 0$)}
    \State $u(k) \gets u^{b}(k)$
\ElsIf{($(E_{mis}(k) \ge 0$) \text{\textbf{AND}} ($P_{mis}(k) < 0$)}
    \State $u_{ac}(k) \gets 0$
    \State $u_{bat}(k) \gets u^{b}_{bat}(k)$
    \State $u_{l}(k) \gets u^{b}_{l}(k)$
\ElsIf{$(E_{mis}(k) < 0)$}
    \If{$(|E_{mis}(k)| \leq u^{b}_{ac}(k) \bar E_{ac})$}
        \State $u_{ac}(k) \gets 0$
        \State $u_{bat}(k) \gets u^{b}_{bat}(k)$
        \State $u_{l}(k) \gets u^{b}_{l}(k)$
    \ElsIf{$(|E_{mis}(k)| \leq u^{b}_{ac}(k) \bar E_{ac}) + \bar E_{l}(k)$}
        \State $u_{ac}(k) \gets 0$
        \State $u_{bat}(k) \gets u^{b}_{bat}(k)$
        \State Compute $u^{ps}_{l}(k)$ using the Priority Stack Controller given in (\ref{eq:PriorityStackLogic}).
        \State $u_{l}(k) \gets u^{ps}_{l}(k)$    
    \EndIf
\EndIf 
\end{algorithmic}
\end{algorithm}
The rule-based controller illustrated in Figure~\ref{fig:RuleBasedSchematic} is developed to check if a mathematically simpler controller based on just sophisticated rules can provide a performance close to the mathematically dense MPC controller. Algorithm~\ref{alg:RuleBasedController} presents the algorithm for the rule-based controller at an arbitrary time index $k$.

This algorithm returns the complete control command $u(k) = [u_{bat}(k),u_{ac}(k),u_{l}(k)]^{T}$, $u_{bat}(k) = [c(k),d(k)]^{T}$ is the battery control command, and $u_{l}(k) = [u_{l_{1}}(k),\dots,u_{l_{n}}(k)]^{T}$ is the other load command. Similarly, $u^{b}(k) = [u^{b}_{bat}(k),u^{b}_{ac}(k),u^{b}_{l}(k)]^{T}$ is the complete control command computed by the Baseline Controller (the Baseline Controller logic is an intrinsic part of this controller), $u^{b}_{bat}(k) = [c^{b}(k),d^{b}(k)]^{T}$ is the battery control command, and $u^{b}_{l}(k) = [u^{b}_{l_{1}}(k),\dots,u^{b}_{l_{n}}(k)]^{T}$ is the other load command. 

The battery charging/discharging dispatch ($E_{bat_{c}}(k)$ and $E_{bat_{d}}(k)$ respectively) are computed as follows;
\begin{align}
    E_{bat_{d}}(k) &= d^{b}(k) min(E_{bat}^{c,dc}, \bar E_{bat}(k)-E_{bat}) . \label{eq:BatteryDispatchDC}
\end{align}
$S_{ac}^{on}(k)$ is a binary indicator for the turning on of the AC (it should be noted that this requires the controller to have a memory of the past on-off state of the AC and this is not shown in Figure~\ref{fig:RuleBasedSchematic} for maintaining simplicity). We compute the energy mismatch of the system ($E_{mis}(k)$) as follows;
\begin{align}
     E_{d}^{b}(k) &= u_{ac}^{b}(k) \bar E_{ac} + \bar E_{l}(k) , \label{eq:Emis1} \\
     E_{a}^{b}(k) &= \bar E_{pv}(k) + E_{bat_{d}}(k) , \label{eq:Emis2} \\
    \bar E_{mis}(k) &= E_{a}^{b}(k) -  E_{d}^{b}(k) , \label{eq:Emis} 
\end{align}
where $E_{d}^{b}(k)$ and $E_{a}^{b}(k)$ are the total house load demand and total available house generation under the action of the baseline controller. Finally, the power mismatch $P_{mis}(k)$ is computed as follows;
\begin{align}
     P_{d}^{b}(k) &= S_{ac}^{on}(k) \bar P_{ac}  , \label{eq:Pmis1} \\
     P_{a}^{b}(k) &= \dfrac{\bar E_{pv}(k)}{\Delta T_{s}} + \bar P_{bat}(k) , \label{eq:Pmis2} \\
    \bar P_{mis}(k) &= P_{a}^{b}(k) - \bar P_{d}^{b}(k) , \label{eq:Pmis} 
\end{align}
where $P_{d}^{b}(k)$ (we do not consider the power demand contribution of the other loads here as the AC startup power is relatively very large, and the AC startup transient time is very small) and $P_{a}^{b}(k)$ are the AC startup power demanded and total available startup power under the action of the baseline controller. It is important to note that both the Baseline and the Rule-Based controllers are reactive as they do not plan in the presence of forecast information.



\section{Simulation Setup}\label{sec:SimulationSetup}
We perform closed-loop simulations of the plant model along with each of the three controllers described in Section~\ref{section:ControlAlgorithms} to compare their performance and generate results. The setup for these simulations is described as follows:

\subsection{Simulation Period and Location}\label{subsec:TimeDuration}
The simulation is carried out for seven days with a time-step of 10 minutes ($\Delta T_{s} = 10$ minutes or $\Delta T_{s} = 1/6$ hours). The period of simulation selected is from Sept. 11, 2017, to Sept. 17, 2017, and the location of the house being simulated is selected to be in Gainesville, FL, USA as hurricane Irma made landfall there on Sept. 11, 2017, and a grid outage was experienced by a large section of its population for a few days.

\subsection{Weather and Load Demand Data}\label{subsec:WeatherLoadData}
We have obtained the weather data (i.e. $\bar{GHI}$, $\bar W_{s}$, and $\bar T_{am}$) for the concerned location and simulation period from the National Solar Radiation Database, see \cite{NSRDBDatabase}. The other load demand data (i.e. $\bar E_{l_{i}}, \; \forall i \in I$ excluding thermal load data) is obtained from Pecan Street Dataport, see \cite{PecanStreetDataport}. We use the load data of a house located in Austin, TX, USA that has both PV and battery installed from the freely available part (has a collection of 75 houses located in Austin in Texas state, California, and New York) of the Pecan Street dataset as it is the closest location to the simulation location (assuming the selected house will be experiencing similar weather to the simulation location). The eight prioritized load classifications as shown in Figure~\ref{fig:PlantSchematic} stem from the various load demands recorded and available in the Pecan Street dataset.

\subsection{Simulation Parameters}\label{subsec:SimulationParameters}
The PV system parameters are as follows: a PV module (Tesla SC325) with a rated power of $0.325 \; kW$ is used to form a $\alpha_{pv} \times 10.075 \; kW$ rooftop PV system, where $\alpha_{pv} \in A_{pv} = \{ 0.25, 0.5, 0.75, 1 \}$ is the PV size factor. The Battery system parameters are as follows: a single Tesla Powerwall is used as the battery system with $\bar E_{bat} = \alpha_{bat} \times 13.5 \; kWh$, $\ubar E_{bat} = 0 \; kWh$ (we allow for complete utilization of storage capacity due to a finite energy system), $\bar P_{bat}^{c,dc} = \alpha_{bat} \times 5 \; kW$ is the rated battery charging/discharging rate, where $\bar E_{bat}^{c,dc} = \bar P_{bat}^{c,dc} \times \Delta T_{s}$, and $\bar P_{bat}^{dc} =  \alpha_{bat} \times 7 \; kW$ is a short time (assumed to be in the order of AC startup transient time) maximum battery discharging rate, where $\alpha_{bat}\in A_{bat} = \{ 0.25, 0.5, 0.75, 1 \}$ is the Battery size factor. It should be noted that we haven't sized the system according to the selected house's load but used the most commercially viable PV-Battery system size ($\alpha_{pv} = 1$ and $\alpha_{bat} = 1$) for a single-family detached house and then progressively reduced the system size to see the effect of resource constraint on the performance of the three controllers. 

The AC system parameters are as follows: $\bar P_{ac}^{rated} = 3 \; kW$ is the rated power of the AC, where $\bar E_{ac} = \bar P_{ac}^{rated} \times \Delta T_{s}$, $\bar T_{h} = 25 \; ^{\circ}C$, $\ubar T_{h} = 23 \; ^{\circ}C$ (the reason for wide dead-band is to conserve energy in a finite energy system), the AC startup power is given as $\bar P_{ac} = (1 - \alpha_{V}) \times \alpha_{I} \times P_{ac}^{rated}$ where $\alpha_{V} = 0.3$ is the AC startup voltage factor which is almost a constant for different AC systems and $\alpha_{I} \in A_{I} = \{ 3,4,5,6,7,8 \}$ is the AC startup current factor (i.e. locked-rotor current factor) which is variable w.r.t. the type of AC system, see \cite{AC:LRA}. Finally, the house thermal system parameters are used from \cite{cui2019hybrid}.

The MPC parameters are as follows: $N = 144$ i.e the planning horizon is of 24 hours with a time-step of 10 minutes, $\lambda_{i} = 1 \; \forall i \in \{ 1,2,3,4,5 \}$, $\bar \Gamma = 1$, and $\ubar \Gamma = -1$. 

\subsection{Computation}\label{subsec:Computation}
The computing system used for performing the simulations is a Windows-based Desktop machine with 32Gb RAM and 3.7 GHz $\times$ 10 CPU. The plant along with the three control algorithms are simulated in MATLAB. The MILP described in (\ref{eq:CostFunction_SingleHouse})-(\ref{eq:Slack_ineq_3}) for the planning horizon of 24 hours ($N=144$) leads to an optimization problem with $576$ binary variables and $1440$ continuous variables.

The optimization solver used to solve the MILP formulated for the MPC controller is GUROBI~\cite{gurobi}. The GUROBI solver parameters that are changed from their defaults are as follows: MIPGap = 1\% (the default value of 0.01\% leads to solver stalling, hence a more pragmatic value is needed given the real-time control requirement), MIPFocus = 2 (the default value is 0 which utilizes a higher solution strategy to strike a balance between finding feasible solutions and proving their optimality, we select the value 2 that focuses on proving optimality as we need the most optimal solution given a finite energy system), Cuts = 2 (the default value is -1 which automatically selects how the cuts should be generated, we select the  value 2 that generates cuts aggressively for a more tightly bound solution), Presolve = 2 (the default value is -1 which automatically selects the presolve strategy, we select the value 2 that aggressively presolves for getting a tighter model before branch and bound can start given the real-time control requirements), and TimeLimit = 500 seconds (the default is no time limit, the value of 500 seconds is smaller than the simulation time-step  of 10 minutes i.e. 600 seconds which helps in the case of a simulation iteration where the solver is stalling and we can still have some suboptimal solution at the end of 500 seconds to provide to the plant and continue on to the next iteration). With these parameters, the solver didn't stall for any of the MPC closed-loop simulations and the average time to solve the MILP was $0.29 \; s$.



\section{Results And Discussion}\label{sec:ResultsDiscussion}


\subsection{Performance Metrics}\label{subsec:PerformanceMetrics}

\subsubsection{Critical Load Resiliency Metric $(LRM_{cri})$}
It measures the amount of critical loads served and is the ratio of total critical loads served to the total critical loads demanded by the system.
\begin{align}
LRM_{cri} &= \frac{\sum\limits^{T_{sim}}_{k=1} E_{cri} (k)}{\sum\limits^{T_{sim}}_{k=1} \bar{E}_{cri} (k)},
\label{eq:LRM_cri}
\end{align}
where $T_{sim}$ is the total number of time steps in a given simulation.

\subsubsection{Other Loads Resiliency Metric $(LRM_{o})$}
It measures the amount of other loads served and is given by the ratio of other loads served to the total amount of other loads over the entire time. 
\begin{align}
 LRM_{o} &= \frac{\sum\limits^{T_{sim}}_{k=1} E_l (k)}{\sum\limits^{T_{sim}}_{k=1} \bar{E}_l (k)},
\label{eq:LRM_o}
\end{align}

\subsubsection {Thermal Resiliency Metric $(TRM_{h})$}
It measures the average thermal comfort and is given by the ratio of the number of times the house temperature is above its upper bound to the total number of simulation time steps.
\begin{align}
 TRM_{h} &= \frac{\sum_{k=1}^{T_{sim}}  \left [ \bar T_{h} - T_{h}(k)  \right]_{+}}{T_{sim}},
\label{eq:TRM_h}
\end{align}

\subsection{Comparison of Controllers for all Simulation cases $ \in A_{I} \times A_{pv} \times A_{bat} $}\label{subsec:ComparisonControllers}
\begin{figure*}[t]
	\centering
	\begin{subfigure}[t]{0.32\textwidth}
		\centering
		\includegraphics[scale=0.195]{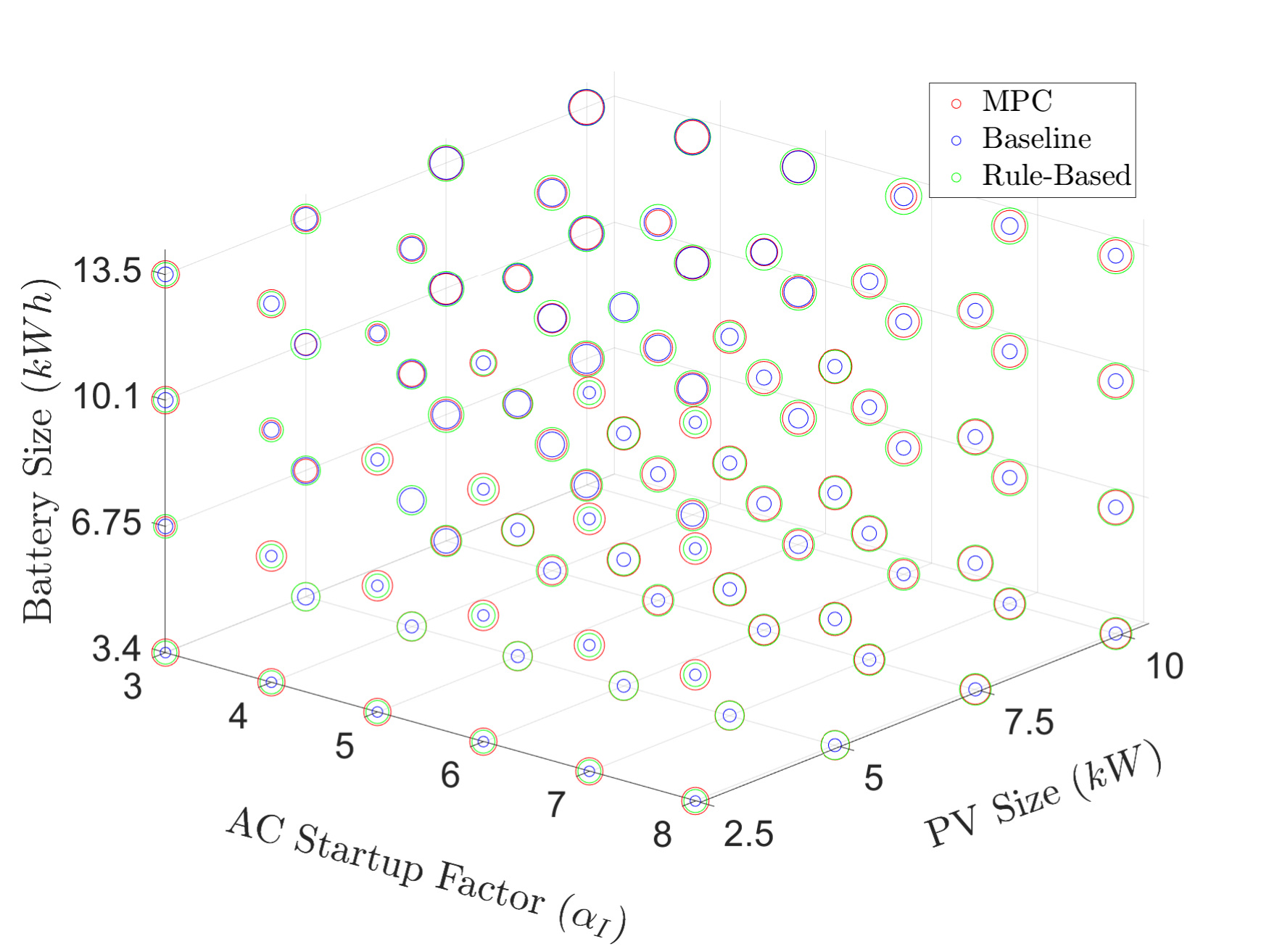}
		\caption{$LRM_{cri}$ Metric.}
		\label{fig:ResultScatter_1}
	\end{subfigure}
	\begin{subfigure}[t]{0.32\textwidth}
		\centering
		\includegraphics[scale=0.195]{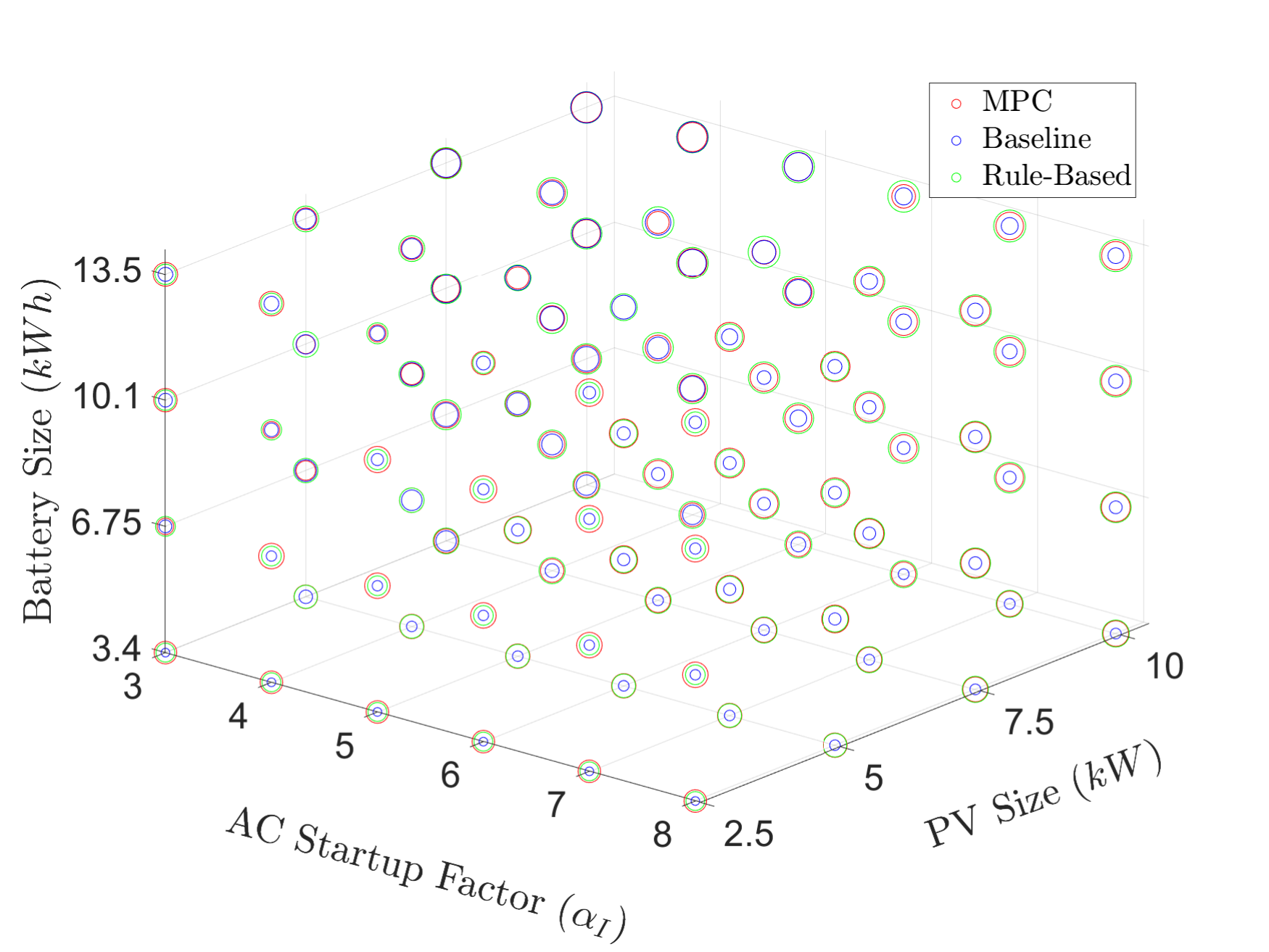}
		\caption{$LRM_{o}$ Metric.}
		\label{fig:ResultScatter_2}
	\end{subfigure}
	\begin{subfigure}[t]{0.32\textwidth}
		\centering
		\includegraphics[scale=0.195]{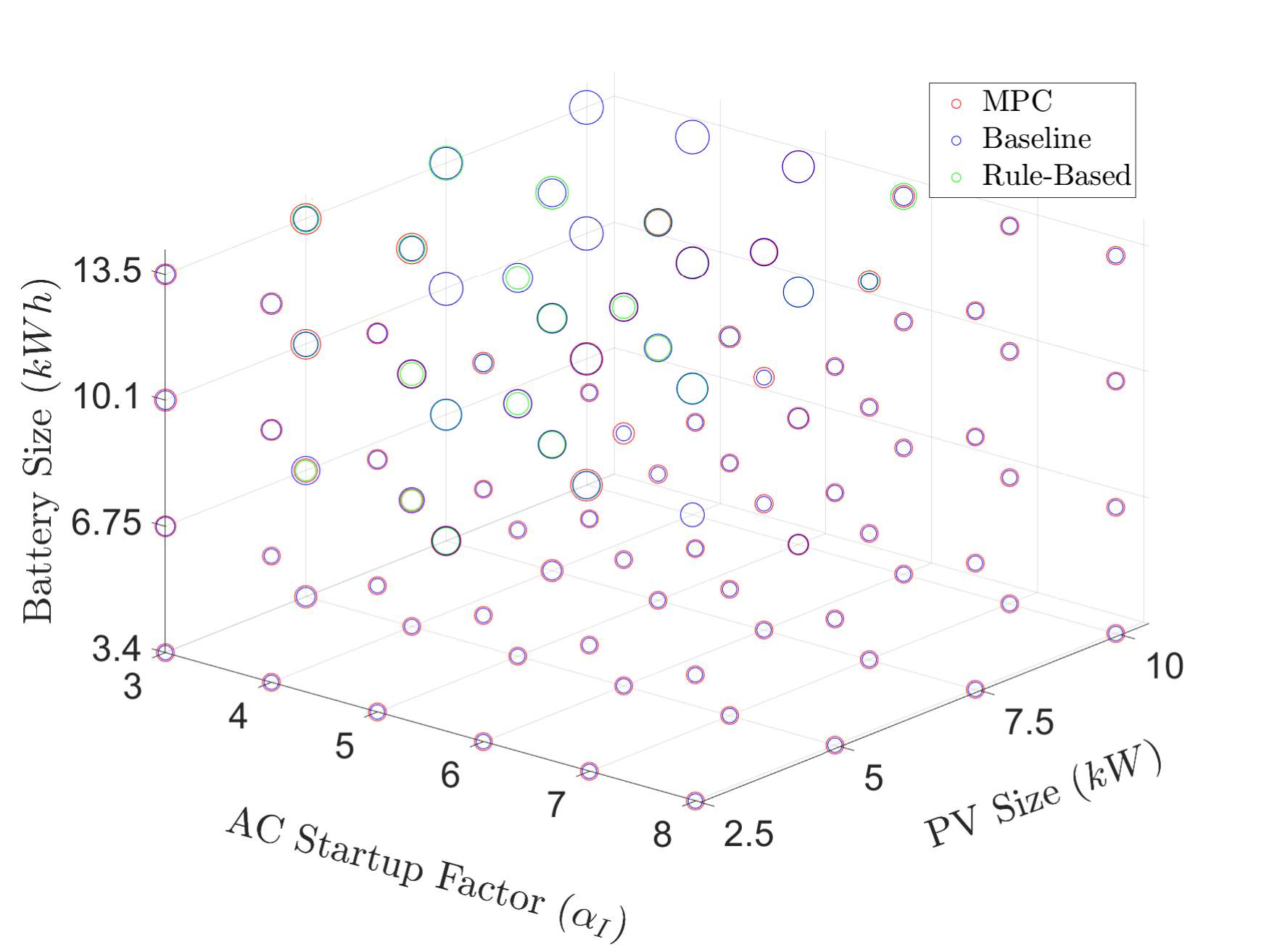}
		\caption{$TRM_{h}$ Metric.}
		\label{fig:ResultScatter_3}
	\end{subfigure} 
	\caption{Performance comparison of the three controllers based on three resiliency metrics, for all the considered simulation cases $ \in A_{I} \times A_{pv} \times A_{bat} $.}
	\label{fig:ResultsScatter}\vspace{-15pt}
\end{figure*}
The Fig.~\ref{fig:ResultsScatter} illustrates the performance comparison of the three controllers (i.e. MPC, Baseline, and Rule-Based) based on the three metrics (i.e. $LRM_{cri}$, $LRM_{o}$, and $TRM_{h}$) for all the considered combinations of the PV size, Battery size, and AC startup current factor (i.e. simulation scenarios). 

The Fig.~\ref{fig:ResultScatter_1} shows the variation of critical load resiliency metric ($LRM_{cri}$) for the three different controllers across all simulation scenarios. It can be observed across all simulation scenarios, that the MPC and Rule-Based controllers perform almost equivalently, and better than the Baseline controller in providing for the critical loads as they are more sophisticated as compared to the Baseline controller. Moreover, the performance of all three controllers improves with an increase in the PV and Battery size and it decreases with an increase in the AC startup factor as expected (excluding the simulation scenarios where the performance metric values saturate due to physical limitations on both the best and worst performance that can be extracted from the system). On the other hand, for a given PV size the performance of the three controllers does not improve significantly for an increase in the Battery size, as an appropriate size of PV is required to generate enough power during the daytime to charge the battery to its capacity. Also, it can be observed that for smaller sizes of PV and Battery the MPC controller performs better than the Rule-Based controller in most scenarios, as its utilization of planning and forecasts within its formulation becomes more important for better performance in more energy-constrained scenarios when compared to only reactive control provided by the Rule-Based controller. 

The Fig.~\ref{fig:ResultScatter_2} shows the variation of other load resiliency metric ($LRM_{o}$) for the three different controllers across all simulation scenarios. The observations that can be made here are equivalent to those made for the critical load resiliency metric, except that the relative numerical values of $LRM_{cri}$ are greater than that of $LRM_{o}$, as critical loads are one of the eight prioritized load components which make up the other load and have the highest priority in the case of  MPC and Rule-Based controllers. 

Fig.~\ref{fig:ResultScatter_3} shows the variation of thermal resiliency metric ($TRM_h$) for the three controllers across all simulation scenarios. It can be observed that the MPC controller can maintain thermal comfort better than the Baseline and Rule-Based controllers in most of the simulation scenarios. This is due to the presence of a competing objective to maintain thermal comfort in the MPC controller's cost function which lends to some relative priority (can be controlled by setting the right $\lambda_{1}$) given to the operation of AC for providing thermal comfort; while no such priority for AC operation is provided in the Rule-Based and Baseline controllers leading to their poor performance. Other observations are equivalent to those made for the critical ($LRM_{cri}$) and other load ($LRM_{o}$) resiliency metrics, with the exception that for some simulation cases the Baseline controller performs better than the MPC and  Rule-Based controllers in providing thermal comfort (one such simulation scenario is explored in detail in the next subsection). Also, it should be noted that the $\lambda_{i}$s of the MPC are not tuned for each of the simulation scenarios individually lending to a more conservative performance.

\subsection{Comparison of Controllers for the Simulation case: $\alpha_{I} = 4, \; \alpha_{pv}=0.5, \; \alpha_{bat}=0.5$}\label{subsec:ComparisonControllersSpecific}
\begin{figure*}[t]
	\centering
    \begin{subfigure}[t]{0.32\textwidth}
		\centering
		\includegraphics[scale=0.39]{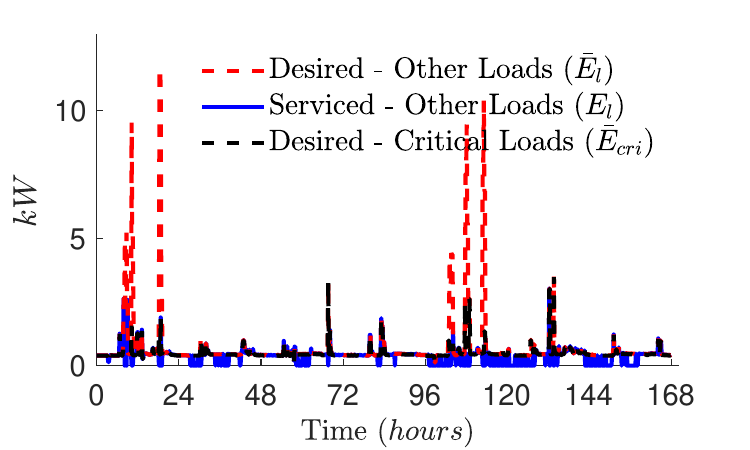}
		\caption{MPC.}
		\label{fig:Result_MPC_load}
	\end{subfigure} 
	\begin{subfigure}[t]{0.32\textwidth}
		\centering
		\includegraphics[scale=0.39]{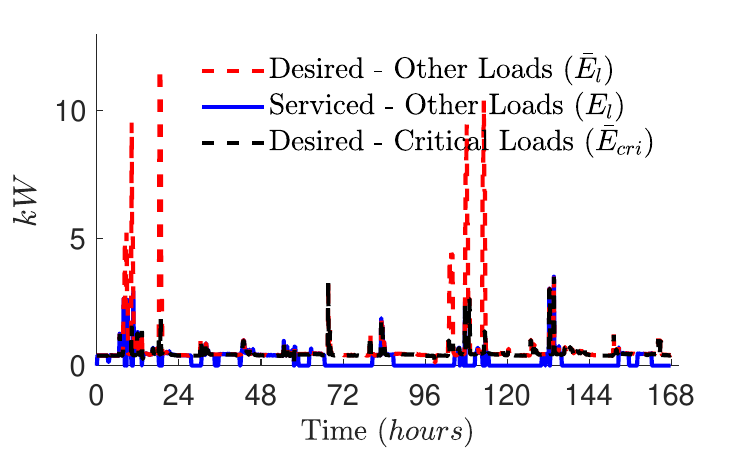}
		\caption{Baseline.}
		\label{fig:Result_Dumb_load}
	\end{subfigure} 
	\begin{subfigure}[t]{0.32\textwidth}
		\centering
		\includegraphics[scale=0.39]{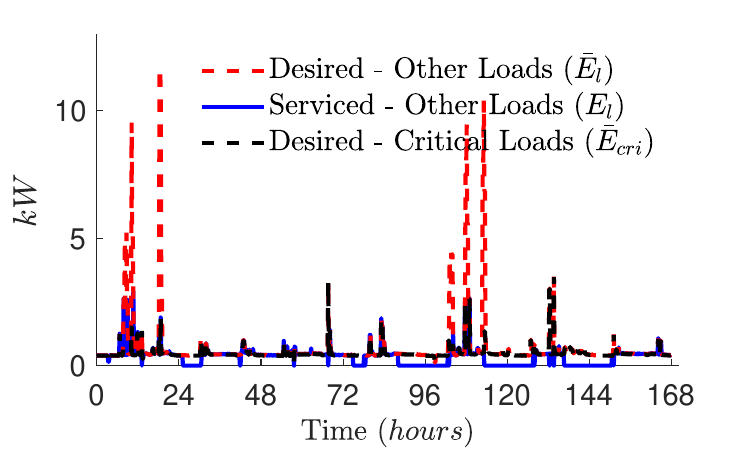}
		\caption{Rule-Based.}
		\label{fig:Result_Smart_load}
	\end{subfigure}

    \begin{subfigure}[t]{0.32\textwidth}
		\centering
		\includegraphics[scale=0.39]{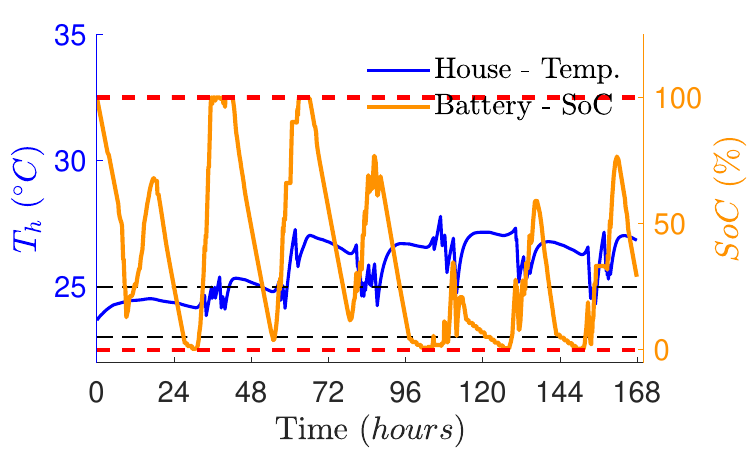}
		\caption{MPC.}
		\label{fig:Result_MPC_Temp}
	\end{subfigure}
	\begin{subfigure}[t]{0.32\textwidth}
		\centering
		\includegraphics[scale=0.39]{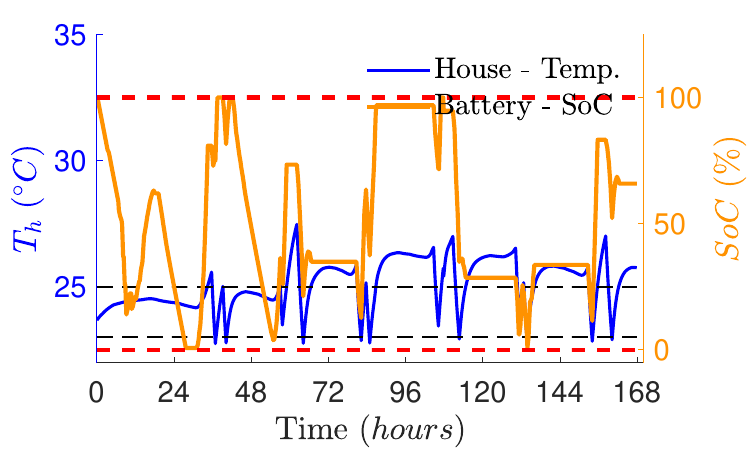}
		\caption{Baseline.}
		\label{fig:Result_Dumb_Temp}
	\end{subfigure}
	\begin{subfigure}[t]{0.32\textwidth}
		\centering
		\includegraphics[scale=0.39]{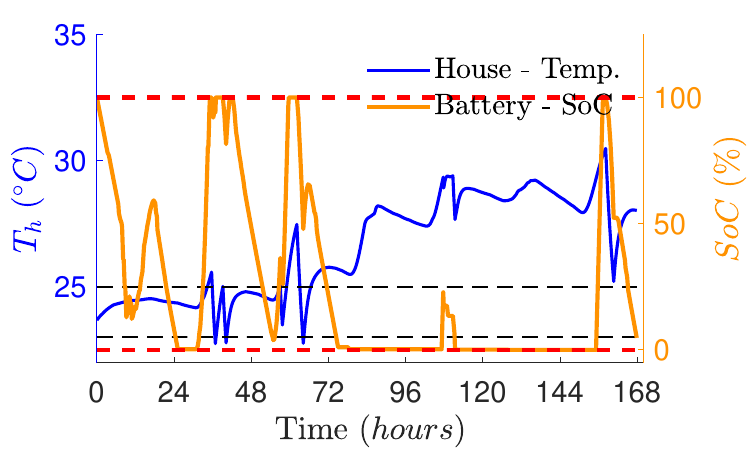}
		\caption{Rule-Based.}
		\label{fig:Result_Smart_Temp}
	\end{subfigure}	
 
    \begin{subfigure}[t]{0.32\textwidth}
		\centering
		\includegraphics[scale=0.39]{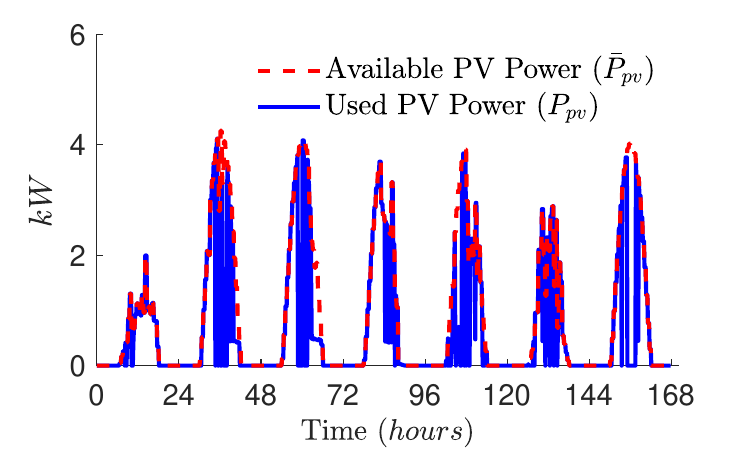}
		\caption{MPC.}
		\label{fig:Result_MPC_PV}
	\end{subfigure} 
	\begin{subfigure}[t]{0.32\textwidth}
		\centering
		\includegraphics[scale=0.39]{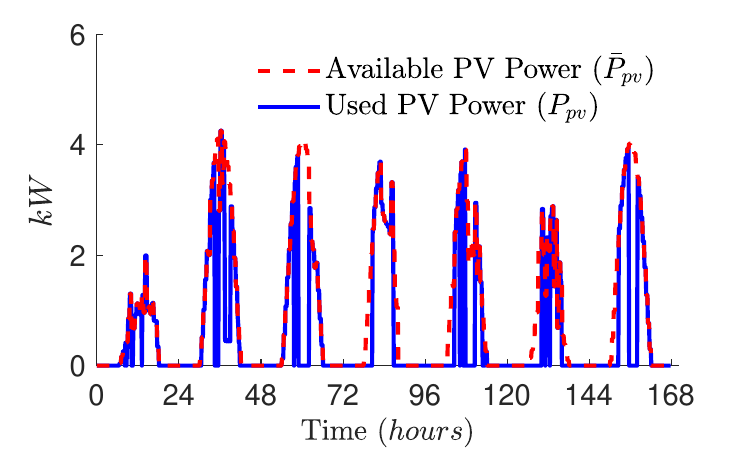}
		\caption{Baseline.}
		\label{fig:Result_Dumb_PV}
	\end{subfigure} 
	\begin{subfigure}[t]{0.32\textwidth}
		\centering
		\includegraphics[scale=0.39]{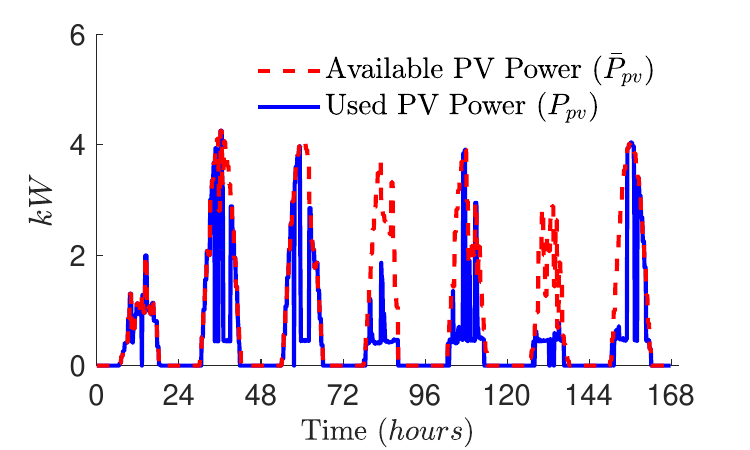}
		\caption{Rule-Based.}
		\label{fig:Result_Smart_PV}
	\end{subfigure} 	
	\caption{Performance comparison of the three controllers for serving the other load along with the critical load (top row), thermal load (middle row), and PV energy utilization (bottom row) for the simulation case: $\alpha_{I} = 4, \alpha_{pv} = 0.5, \alpha_{bat} = 0.5.$.}
	\label{fig:Results_graph}\vspace{-15pt}
\end{figure*}

Fig.~\ref{fig:Result_MPC_load}, Fig.~\ref{fig:Result_Dumb_load} and Fig.~\ref{fig:Result_Smart_load} illustrate the time series plots of the desired and serviced loads due to MPC, Baseline, and Rule-Based controllers respectively for the entire simulation period. It can be observed that all three controllers can provide for at most the desired critical load as the size of the PV/Battery system is not adequate to provide for all of the desired load; except for the initial simulation period where they provide for the desired load greater than the critical load as the house temperature is within bounds (turning on of the AC is not required) and the battery state-of-charge (SoC) is high (enough stored energy). Table~\ref{tab:metrics_table} shows that the MPC and Rule-Based controllers outperform the Baseline controller in providing for the desired loads, this is expected as the Baseline controller is not as sophisticated in load management as the other two controllers. Furthermore, the MPC and Rule-Based controllers perform equivalently in providing for the other loads, but the MPC controller is slightly worse in providing for the critical loads as it prioritizes the thermal load to some extent as opposed to the Rule-Based controller which completely sheds the thermal load in favor of critical/other loads in an energy-constrained situation.  

\begin{table}[htpb]
\centering
\caption{Performance metrics for the simulation case: $\alpha_{I} = 4, \alpha_{pv} = 0.5, \alpha_{bat} = 0.5.$}
\begin{tabular}{|l|c|c|c|}
\hline
                    & \multicolumn{1}{l|}{\textbf{$LRM_{cri}$}} & \multicolumn{1}{l|}{\textbf{$LRM_o$}} & \multicolumn{1}{l|}{\textbf{$TRM_h$}} \\ \hline
\textbf{MPC}        & 0.65                                      & 0.53                                  & 0.46                                  \\ \hline
\textbf{Baseline}   & 0.45                                      & 0.37                                  & 0.55                                  \\ \hline
\textbf{Rule-Based} & 0.66                                      & 0.53                                  & 0.36                                  \\ \hline
\end{tabular}
\label{tab:metrics_table}
\end{table}

Fig.~\ref{fig:Result_MPC_Temp}, Fig.~\ref{fig:Result_Dumb_Temp} and Fig.~\ref{fig:Result_Smart_Temp} illustrate the time series plots of the house temperature and battery SoC due to MPC, Baseline, and Rule-Based controllers respectively for the entire simulation period. From these time series plots along with the metric values presented in Table~\ref{tab:metrics_table} it can be observed that the Baseline controller outperforms the other two controllers in providing thermal comfort, but due to lack of any energy management logic, it prioritizes the thermal and other loads equally and in this simulation scenario the thermal load turns out as the winner. On the other hand, the MPC controller performs significantly better than the Rule-Based controller in providing thermal comfort and maintaining battery SoC as it has a mathematical formulation to prioritize both, while the Rule-Based controller lacks such a formulation leading to a dead battery (approximately between 72 to 150 hours) which leads to no power availability for AC startup (still the critical/other loads are being served). Around the 72, 96, 120, and 144-hour marks where the temperature is out of bound for both the MPC and Baseline controllers, it can be observed that in the case of MPC, it is a planning-based decision in the presence of forecast information to not serve the thermal load and instead prioritize the charging of the battery and serving of the other load; while in the case of Baseline controller the thermal load along with the other load is not served as trying to turn on the AC causes a power limit violation leading to no power transfer in the system. Fig.~\ref{fig:Result_MPC_PV}, Fig.~\ref{fig:Result_Dumb_PV} and Fig.~\ref{fig:Result_Smart_PV} illustrate the time series plots of the PV energy utilization against the available PV energy for  MPC, Baseline, and Rule-Based controllers respectively for the entire simulation period, these time series plots corroborate the previous observations.

Hence, from all the results and the associated discussion, it is clear that the MPC controller with its capability for intelligent planning in the presence of forecast information is well-suited for providing energy resiliency to a house during an unplanned outage caused by extreme weather.


\section{Conclusion}\label{sec:Conclusion}
We present an MPC-based controller for providing energy resiliency to a house equipped with roof-top solar PV and battery energy storage during unplanned outages caused by extreme weather events while maintaining thermal comfort through intelligent AC operation. We compare our proposed MPC controller with a Baseline and a Rule-Based controller based on three resiliency metrics. The simulation results show that generally, the performance of the three controllers is directly proportional to the PV and battery size while being inversely proportional to the AC startup factor. The MPC controller performs better than the other controllers in the more energy-constrained scenarios (smaller PV-battery size, larger AC startup factor) in providing both thermal comfort and servicing critical/other loads in a balanced manner. This study illustrates that a controller based on intelligent planning and forecast information utilization is more reliable than reactive rule-based controllers in providing home energy resiliency against extreme weather events. 

In the future, we plan to explore: the performance of the MPC controller during extreme cold events, the sensitivity of the MPC controller to forecast errors, the sizing of PV and battery based on resiliency requirements, the development of a RL based controller to avoid solving the MILP in the MPC controller, and the development of a distributed control framework to provide home energy resilience against extreme weather events for a community of heterogeneous (can have different distributed energy resources) houses.



\section*{Acknowledgment}
This work is funded by the NSF award 2208783.


\bibliographystyle{IEEEtran}
\bibliography{./BibFile}


\end{document}